\documentclass[10pt]{article}  
\pdfoutput=1
\usepackage{times}
\usepackage{epsf,graphicx,multicol,float}
\usepackage{tocloft}
\usepackage{natbib,citesort}

\usepackage{color} 

\begin{document}

{\LARGE \textbf{Physical Properties of Biological Membranes}}\\
\begin{tabular}{p{16cm}}
    \\ \hline\\
\end{tabular}
Thomas Heimburg, \textit{Niels Bohr Institute,}\textit{University of Copenhagen, Copenhagen, Denmark}

\begin{multicols}{2}[]
    \tableofcontents
\end{multicols}

\begin{tabular}{p{16cm}}
    \\ \hline\\
\end{tabular}

\begin{multicols}{2}[]

\section{Introduction}
\label{sec1}

\begin{figure*}[htb!]
    \begin{center}
	\includegraphics[width=12cm]{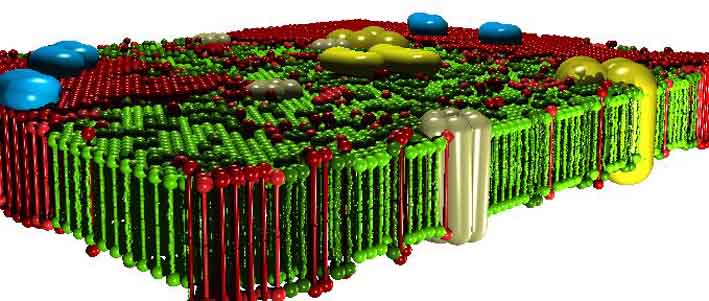}
	\parbox[c]{14cm}{ \caption{\textit{Schematic representation 
	of a biological membrane with various lipids in different 
	physical states and embedded or adsorbed proteins. Courtesy 
	Heiko Seeger, NBI.}
	\label{Figure1A}}}
    \end{center}
\end{figure*}

Biological membranes are thin layers surrounding cells and their organelles, which define inside and outside of compartments.  Their major components are lipids and proteins.  Lipids are small amphiphilic molecules that are hydrophilic on one side and hydrophobic on the other side.  When brought into contact with water most lipids spontaneously form bilayer membranes.  Our present picture of biomembranes stems from \cite{Gorter1925}, \cite{Danielli1935} and  \cite{Singer1972}.  The latter paper described the membrane as a two-dimensional bilayer of fluid lipids with proteins adsorbed or embedded.  The Singer-Nicolson paper still dominates our present understanding of membranes, but it experienced some modifications. \cite{Mouritsen1984} proposed that the membrane is laterally heterogeneous as schematically shown in Fig. \ref{Figure1A}.  Here one sees domain formation and proteins that associate with different domains with different likelihood.\\
The typical thickness of a lipid bilayer is about 5nm.  Thus, these are extremely thin layers as compared to the typical dimensions of cells (several 10 $\mu$m).  Such membranes are flexible, heterogenous and permeable.  However, these properties depend on the conditions, for example on temperature, pressure, electrical field, pH, salt concentration, but also the presence of proteins and protein conformation.  This basically means that the state of a biological membrane depends on all thermodynamic variables.  These dependencies are discussed throughout this article.  Thus, in this article we focus rather on the properties of membranes as a whole than on the properties and functions of the individual molecules.\\
The composition of membranes is complex (see section \ref{sec2} for details).  There are hundreds or evens thousands of different lipid species and further thousands of different membrane proteins.  The composition of membranes is different not only between different species, but also between different cell types of the same organism, and even between the membranes of different organelles within the same cell.  Proteins act, e. g., as catalysts and the importance of their role seems intuitively clear, while the role of the lipid membrane and the range of different lipid compositions is less obvious and there is no agreement in the  biophysical community yet on what the purpose of the heterogeneity of  the lipids precisely is.  \\
Lipid membranes can undergo transitions from a rigid into a liquid state (section \ref{sec3}) where both the lateral order of lipid molecules and the order of the lipid hydrocarbon chains changes.  In biological membranes that occurs at temperatures that typically are slightly below body or growth temperature.  Both states form two-dimensional membranes but their physical properties may be quite different.  The lipid composition has an influence on membrane melting. The melting events in membranes lead to changes in the distribution of lipids within the membrane plane.  In experiments, one finds regions of different composition and state.  If they are of macroscopic dimensions they are called phases, if they are on microscopic or nanoscopic scales they are called domains or in cell biology sometimes `rafts'.  The investigation of domain formation received much interest because of its putative influence on reaction cascades between proteins, and on diffusion processes within membranes.  The melting transition in membranes is also interesting because of its influence on the elastic constants of membranes.  In particular, at their melting temperatures the membranes are very soft and easy to bend (section \ref{sec5}).  Related to this, the membranes also become very permeable for water, ions and larger molecules (section \ref{sec7}).  Under some conditions, they may even propagate mechanical perturbations or pulses that resemble nerve pulses (section \ref{sec8}).

\section{Composition of membranes}
\label{sec2}
When talking about the composition of biological membranes one has to distinguish proteins that are encoded in the  genome, and the lipid composition that is not encoded in the genome. The latter rather adapts to the environmental conditions, partially by the control of membrane active proteins that display an activity depending on the physical state of the membrane.
\begin{figure}[H]
\hspace*{0cm}
\includegraphics[width=7.5cm]{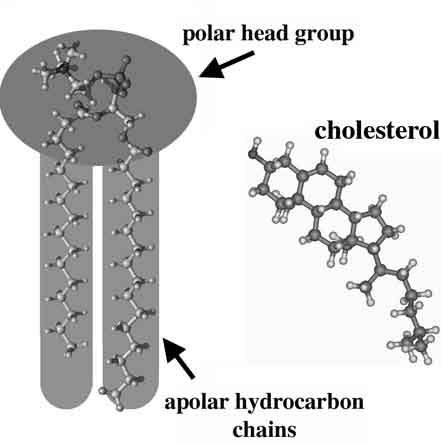}\\
\hspace*{0.5cm}\parbox[c]{7cm}{\caption{\textit{Left: Dipalmitoyl phosphatidylcholine (DPPC) is a phospholipid molecule with zwitterionic polar head group and apolar fatty acid chains. Right: The steroid lipid cholesterol displays a quite different and much more rigid structure.  }
\label{Figure2A}}}
\end{figure}

\subsection{Lipid composition}
\label{sec2.1}
There are hundreds of different lipids.  It is only loosely defined what precisely counts as a lipid.  What lipids have in common is that they are small amphiphilic molecules that dissolve in membranes.  By far the majority of these lipids are phospholipids with two hydrocarbon chains and a phosphate containing head group as shown in Fig.  \ref{Figure2A} (left).  However, one also finds minor fractions of lipids that only have one chain, or molecules of completely different structure as cholesterol (Fig.  \ref{Figure2A}, right). This is a quite rigid molecule not forming membranes on its own. It is nevertheless a major component in some biological membranes, e.g. in erythrocytes \cite[]{Sackmann1995a}. One of the most obvious features being controlled by the chemical nature of the lipids is the melting point of the membranes that is discussed in more detail in section \ref{sec3}.  Another property is the electrostatic potential of the membranes if the lipid head groups are charged.

\subsubsection{Hydrocarbon chains and chain length}
\label{sec2.1.1}
Most lipids possess two hydrocarbon chains as shown in Fig.  \ref{Figure2A}. These chains vary in length from 12 to about 24 carbons. A second feature is the degree of unsaturation in the chain, i.e. the number of double bonds in the chain. The chemical nature of the chains has a major influence on the melting temperatures of membranes as described in section \ref{sec3}.

\subsubsection{Head groups}
\label{sec2.1.2}
Lipids may also possess a variety of different polar head groups.  Most lipids are phospholipids containing a phosphate group.  About 40 mol\% of the lipids in eucariotic cells are phosphatidylcholines (= lecithins) which are zwitterionic. They carries one negative and one positive charge in the physiological pH range. One example is shown in Fig. \ref{Figure2A}. Another abundant heat group is the zwitterionic phosphatidylethanolamine that also contributes about 40 mol\%.  The rest are various other lipids that occur in smaller fractions, e.g. the negatively charged phosphatidylserines, phosphatidylglycerols, phosphatidylinositols and cardiolipins. Other lipids are sphingolipids and ceramides.\\

\subsection{Lipid composition of biomembranes}
\label{sec2.2}
The lipid composition of membranes from different sources can be quite different. Examples are shown in Table \ref{Table2.1}.  Human erythrocytes and rat liver plasma membranes, for instance, contain about 20 weight\% cholesterol while rat liver mitochondria contain only 3 weight\%.  The phosphatidylcholine content of the endoplasmic reticulum is 48 weight\% while the content in nerve myelin is only 11 weight\%.  Other examples can be found in \cite{Heimburg2007a} or \cite{Sackmann1995a}.  The differences in lipid composition between different cells, and even the organelles of the same cell are remarkable, but the reason is not completely understood.
\setlength{\tabcolsep}{0.17 cm}
\begin{table*}[t]     
    \parbox[c]{14cm}{\caption{\textit{Head group composition of the
    membranes of some mammalian liver cells,\index{liver cells}
    erythrocytes \index{erythrocytes}, and nerve cells\index{neuron}
    in weight percent.  Adapted from \cite{Jamieson1977}.
    Abbreviations: PC = phosphatidylcholines, PE =
    phosphatidylethanolamines, PS = phosphatidylserines, PI =
    phosphatidylinositols, SM = sphingomyelin, CL = cardiolipin.}}}
    \setlength{\tabcolsep}{4pt}
    \hspace*{1.0cm}
    \begin{tabular}{@{}l|*{9}{c}@{}}
	    \hline
	    {\small Membrane} & {\small PC} & {\small PE} & {\small
	    PS} & {\small PI} & {\small SM} & {\small CL} &
	    {\small Glycolipid} &{\small Cholesterol} & {\small Others}\\
	    \hline
	    {\small Erythrocyte (human)} & {\small 20} & {\small 18} & {\small
	    7} & {\small 3} & {\small 18} & {\small --} &
	    {\small 3} &{\small 20} & {\small 11}\\
	    {\small Plasma (rat liver)} & {\small 18} & {\small 12} & {\small
	    7} & {\small 3} & {\small 12} & {\small --} &
	    {\small 8} &{\small 19} & {\small 21}\\
	    {\small ER} & {\small 48} & {\small 19} & {\small
	    4} & {\small 8} & {\small 5} & {\small --} &
	    {\small tr} &{\small 6} & {\small 10}\\
	    {\small Golgi} & {\small 25} & {\small 9} & {\small
	    3} & {\small 5} & {\small 7} & {\small --} &
	    {\small 0} &{\small 8} & {\small 43}\\
	    {\small Lysosome} & {\small 23} & {\small 13} & {\small
	    --} & {\small 6} & {\small 23} & {\small $\approx$ 5} &
	    {\small --} &{\small 14} & {\small 16}\\
	    {\small Nuclear membrane} & {\small 44} & {\small 17} & {\small
	    4} & {\small 6} & {\small 3} & {\small 1} &
	    {\small tr} &{\small 10} & {\small 15}\\
	    {\small Mitochondria} & {\small 38} & {\small 29} & {\small
	    0} & {\small 3} & {\small 0} & {\small 14} &
	    {\small tr} &{\small 3} & {\small 13}\\
	    {\small Neurons} & {\small 48} & {\small 21} & {\small
	    5} & {\small 7} & {\small 4} & {\small --} &
	    {\small 3} &{\small 11} & {\small 1}\\
	    {\small Myelin} & {\small 11} & {\small 17} & {\small
	    9} & {\small 1} & {\small 8} & {\small --} &
	    {\small 20} &{\small 28} & {\small 6}\\
	    \hline
    \end{tabular}
    
    \label{Table2.1}
\end{table*}

\subsection{Dependence on thermodynamic variables}
\label{sec2.3}
It seems likely that the control of the thermodynamic parameters associated to the melting points of membranes (described in more detail in section \ref{sec3}) belong to the most important features that are influenced by the lipid composition. This is a result of studying the lipid compositions of various organisms that live at conditions of variable temperature and pressure.

\subsubsection{Growth temperature}
\label{sec2.3.1}
It has been shown that the lipid composition of cold-blooded organisms depends strongly on the ambient growth temperature.  Bacillus subtilis bacteria show a largely altered lipid composition when grown at different temperatures \cite[]{vandeVossenberg1999}.  The composition of trout livers are very different in trouts caught in winter (water temperature $\approx$ 5$^{\circ}$C) and trouts from summer (water temperature $\approx$ 20$^{\circ}$C) \cite[]{Hazel1979}.  In particular, the amount of unsaturated lipids was much higher in the livers of trouts from winter while the amount of saturated lipid was higher in summer.  Unsaturated lipids have much lower melting temperatures than saturated lipids.  It seems as if organisms adapt their lipid compositions according to temperature in order to maintain a particular physical state of their membranes.

\subsubsection{Growth pressure}
\label{sec2.3.2}
\cite{deLong1985} investigated barophilic marine bacteria.  They found that the lipid composition of their membranes was highly dependent in the growth pressure.  As temperature, pressure also has an influence on membrane melting. At high pressure the melting point increases.  \cite{deLong1985} found that the ratio of saturated to unsaturated chains increased linearly with ambient pressure.

\subsubsection{Solvents and anesthetics}
\label{sec2.3.3}
Similar effects have been seen when \textit{E. coli} bacteria are grown in media containing solvents as acetone or aniline \cite[]{Ingram1977}. Depending on the  concentration of the solvents one observes drastic changes in the  composition of the membranes. In section \ref{sec3.1.3} it is shown that the adaptation of the lipid composition is accompanied by a change of the melting temperature of membranes.

\subsection{Membrane proteins}
\label{sec2.4}
\begin{figure}[H]
    \hspace*{0.5cm}
    \includegraphics[width=7cm]{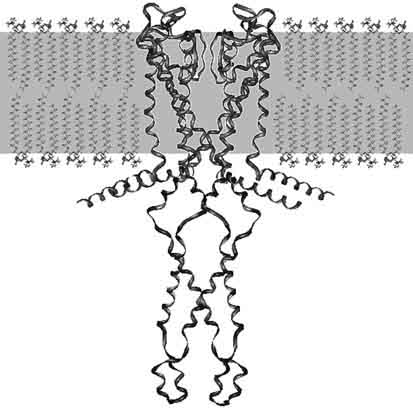}\\
    \hspace*{0.5cm}\parbox[c]{7cm}{\caption{\textit{The potassium 
    channel protein spans through the lipid bilayer. Crystal 
    structure of the protein from \cite{Cortes2001}.}
    \label{Figure2B}}}
\end{figure}

In contrast to the lipid structures proteins are genetically encoded. The human genome contains about 30.000 genes that encode proteins, many of them being membrane proteins.  Membrane proteins are much more difficult to crystallize than soluble proteins.  Therefore the number of known membrane protein structures is still small.  Membrane proteins have various putative functions, but many are thought to be involved in transport processes across membranes.  This includes the active transport of protons by the light driven rhodopsin, of sodium and potassium by the ATP-consuming Na$^+$/K$^+$-ATPase, and passive transport by ion channel proteins, e.g., Na$^+$-channel and K$^+$-channel (Fig.\ref{Figure2B}).\\
In this article we focus on the physical properties of membranes as a whole rather than on the function of their components.  The interested reader may therefore refer to other excellent publications on membrane proteins.

\section{Membrane melting}
\label{sec3}
One of the important features of membranes is that they can melt. Melting processes are order transitions between two planar membrane phases or states that display different physical properties.  

\subsection{Chain melting}
\label{sec3.1}
The molecular origin of chain melting is shown in Fig. \ref{Figure3A}. At low temperatures, lipid chains are ordered into an all-trans configuration (Fig. \ref{Figure3A}, top left). At high temperatures these chains are disordered due to rotations around the C-C bonds within the lipid chains (Fig. \ref{Figure3A}, top right). The membranes are in an ordered 'gel' phase at low temperature while they are in a disordered 'fluid phase' at high temperatures  (Fig. \ref{Figure3A}, center). The detailed nature of the phases involved in the melting process is described in more detail in section \ref{sec4}.

\subsubsection{Temperature dependence and heat capacity}
\label{sec3.1.1}
\begin{figure}[H]
    \hspace*{1.0cm}
    \includegraphics[width=5.5cm]{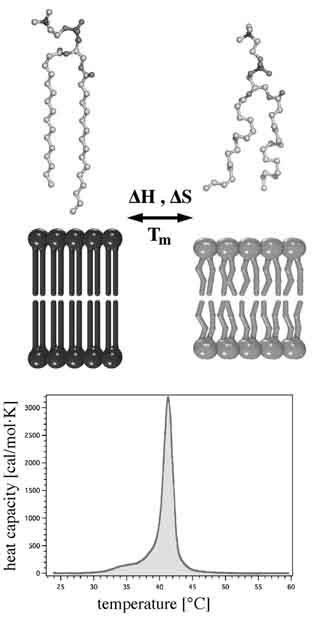}\\
    \hspace*{0.5cm}\parbox[c]{7cm}{\caption{\textit{Upon temperature 
    increase, lipids undergo a melting transition where chains become 
    unordered. This melting can be observed as a peak maximum of the 
    heat capacity as measured in calorimetry.}
    \label{Figure3A}}}
\end{figure}

In the chain melting process, the molecules absorb heat (enthalpy) and the entropy increases due to the increase in the number of possible chain configurations. The melting processes is cooperative, meaning that the lipids do not melt independently of each other. The absorption of heat is typically monitored by measuring the heat capacity, that is the amount of heat absorbed per degree temperature increase. The heat capacity at constant pressure is defined by $c_p=(dH/dT)_p$. It is shown in Fig. \ref{Figure3A} (bottom). At the melting temperature, $T_m$ it displays a pronounced maximum.

\subsubsection{Pressure dependence}
\label{sec3.1.2}
During the melting process membranes also change their volume and area \cite[]{Heimburg1998}. As a rule of thumb the volume changes by about 4\% and the area by about 25\%. Due to these changes, the melting temperature is pressure dependent. If one increases the hydrostatic pressure by about 40 bars the melting temperature increases by 1 degree. Similarly if one increases the lateral tension within the membrane, the melting temperature also changes. The latter effect can be studied on a Langmuir trough where lipid monolayers are compressed by changing the film area while simultaneously monitoring the lateral pressure\cite[]{Mohwald1990}. 

\subsubsection{Solvent dependence}
\label{sec3.1.3}
There are other factor that change the melting temperature, for example the partitioning of organic solvents in membranes. Typically, solvents don't dissolve well in the ordered low temperature phase but dissolve quite well in the fluid phase. This gives rise to a lowering of the melting temperatures due to a well known effect called freezing point depression. As discussed in section \ref{sec9} this is likely to be related to the anesthesia since most anesthetics resemble organic solvents.

\subsubsection{Voltage dependence}
\label{sec3.1.4}
Membranes contain zwitterionic and charged lipids. The zwitterionic lipids have a head group that is a electric dipole. Thus, a monolayer of lipids possesses a strong dipole potential across that layer that can be of the order of 500 mV (e.g., \cite[]{Pickard1979}). Upon area changes in a monolayer experiment one finds that the lateral pressure changes in close relationship to this dipole potential. This implies that similarly to a change in pressure, voltage changes can induce a transition in the lipid monolayers. Since a membrane consists of two monolayers, these potentials partially compensate. In the presence of charged lipids the membranes additionally possess a net electrostatic potential that is discussed in section \ref{sec6}.

\subsubsection{Dependence on membrane proteins}
\label{sec3.1.5}
As already mentioned in section \ref{sec2.4}, many proteins are embedded into or absorbed on membranes. This also has an influence on melting processes \cite[]{Zhang1995b}. Some proteins shift transitions to higher temperatures, e.g., the band 3 protein of erythrocytes \cite[]{Morrow1986} while others shift it rather to lower temperatures, e.g., cytochrome b5 \cite[]{Freire1983}. For integral proteins, the shift is related to the hydrophobic matching between proteins and lipid bilayer thickness \cite[]{Mouritsen1984}.

\subsection{Melting in biomembranes}
\label{sec3.2}
\begin{figure}[H]
    \hspace*{0.5cm}
    \includegraphics[width=6.5cm]{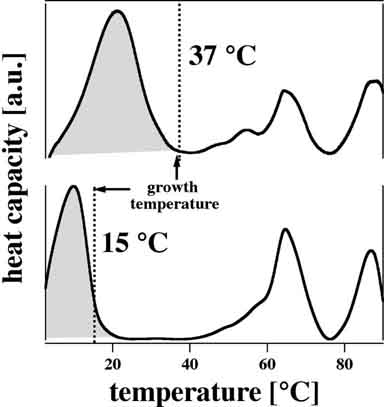}\\
    \hspace*{0.5cm}\parbox[c]{7cm}{\caption{\textit{The native cell
    membranes of E.coli bacteria also display chain melting of their
    membranes at temperature slightly below growth temperature (shaded
    grey).  The lipid membrane melting peaks adapt to the growth
    temperature (dotted line).  The peaks above growth temperature are
    protein unfolding peaks.  Their position does not depend on growth
    temperature.}
    \label{Figure3B}}}
\end{figure}

Most of the above has been investigated in model membranes or monolayers made of synthetic lipids. However, all of it gains biological relevance due to the fact that biological membranes also possess such melting transitions of their membranes, typically slightly below body temperature. This can be seen in Fig. \ref{Figure3B}. Here, heat capacity profiles of native \textit{E.coli} membranes are shown. The two traces are from bacteria grown at two different temperatures, one at 37$^{\circ}$C and the other one at 15$^{\circ}$C, respectively. A number of peaks can be seen in these traces. The membrane melting peaks are shaded in grey. The other peaks represent the unfolding of proteins at higher temperatures. The membrane melting peak occurs at lower temperature for bacteria grown at 15$^{\circ}$C than for those grown at 37$^{\circ}$C. The bacteria adapt their membrane lipid compositions such that the melting point of the membranes is always below growth temperature. That the lipid composition responds to the changes in the environment has already be described in section \ref{sec2}.

\section{Phases and domains}
\label{sec4}

\begin{figure}[H]
    \hspace*{0.0cm}
    \includegraphics[width=7.5cm]{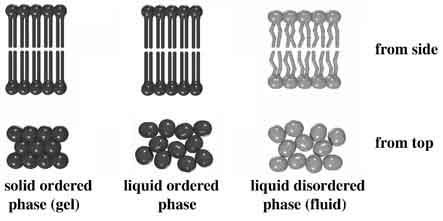}\\
    \hspace*{0.5cm}\parbox[c]{7cm}{\caption{\textit{The three most
    common membrane phases seen as cross sections and from top: solid
    ordered (SO), liquid ordered (LO) and liquid disordered (LD)
    phases.  The first two phases (SO and LO) are low enthalpy phases,
    while the LD phase is a high enthalpy phase mostly existing at
    high temperatures.  }
    \label{Figure4A}}}
\end{figure}

The thermodynamics of ensembles of molecules allows for the formation  of quite different phases. If the conditions are right more than one phase may coexist. This happens in temperature, pressure or  concentration intervals well defined by the thermodynamic parameters of the system. The graphical representation of the coexistence of such phases is called a `phase diagram' (cf. Fig. \ref{Figure4B}).\\
The most common phases found in artificial and biological membranes are the solid ordered (SO) phase, the liquid ordered (LO) phase and the liquid disordered (LD) phase (Fig.  \ref{Figure4A}).  'Solid' means that the lipids arrange on a (typically triangular) lattice in the membrane plane while 'liquid' indicates the loss of lateral packing (Fig.  \ref{Figure4A}, bottom).  The attributes 'ordered' and 'disordered' indicate the internal order of the lipid chains (Fig.  \ref{Figure4A}, top).  The transition between ordered and disordered states is linked to the uptake of significant amounts of enthalpy (i.e., heat), as already described in section \ref{sec3} (Fig.  \ref{Figure3A}).  The SO phase is also often called the `gel phase' and the trivial name of the LD phase is the `fluid phase'.  In the literature one often finds those trivial names of the phases.\\
Typically, the different phases differ in their physical properties.  For instance, they may display very different elasticities, auto-diffusion constants or electrostatic potential. Under most circumstances biological membranes seem to exist in the LD phase. Due to the proximity to melting transitions, however, this can be fine tuned by the organism.

\subsection{GibbsÕ phase rule}
\label{sec4.1}
In a lipid mixture like the biological membrane one may find more then the three phases mentioned above.  For instance, one could have two LD phases coexisting. \\
Let us assume we have $K$ lipid (or protein) components, and $P$ phases.  Of course, one has always at least one phase, i.e. $P \ge 1$. Further, we introduce the number of degrees of freedom, $F$, that indicates how many variables can be changed independently of each other. Such variables are the concentrations of the different membrane component and the temperature.  The number of degrees of freedom must always be positive, i.e. $F \ge 0$.\\

\begin{figure}[H]
    \hspace*{0.0cm}
    \includegraphics[width=7.5cm]{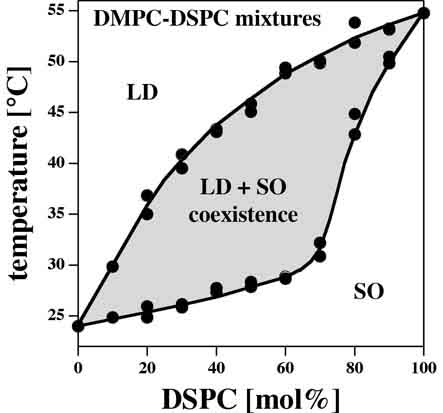}\\
    \hspace*{0.5cm}\parbox[c]{7cm}{\caption{\textit{Phase diagram of 
    a simple binary lipid mixture. The grey-shaded region indicates 
    the regime of temperature and concentration where a LD (gel) and 
    a LD (fluid) phase coexist. Adapted from \cite{Hac2005}.}
    \label{Figure4B}}}
\end{figure}

There is a famous thermodynamic rule about how many  different phases can coexist for a given number of components and  degrees of freedom that is called `Gibbs' phase rule'. At constant pressure it is given  by:
\begin{equation}
    F=K-P+1\qquad\mbox{Gibbs' phase rule}
    \label{eq:4.1}
\end{equation}
Let us give a simple example for this.  In a binary lipid mixture as described in Fig.  \ref{Figure4B} one has two components, i.e. $K=2$. Let us now check all possibilities:
\begin{eqnarray}
    P=1 & \rightarrow & F=2 \nonumber\\
    P=2 & \rightarrow & F=1 \\
    P=3 & \rightarrow & F=0  \nonumber
    \label{eq:4.2}
\end{eqnarray}
The first case indicates that one has only one phase, and one can vary both the concentration of the second component and the temperature freely without leaving the one-phase region. The second  case indicates that one may have two different phases coexisting. However, when one of the variables (for example temperature) is varied, the second variable (concentration of the components in the two phases) is automatically determined and can not be varied freely. The last case in eq. \ref{eq:4.2} tells us that there may be points (with both defined temperature and composition) where three phases may coexist.\\

\subsection{Phase diagrams}
\label{sec4.2}
An example for a simple phase diagram is shown in Fig.\ref{Figure4B}. It represents the mixture of two synthetic lipids dimyristoyl phosphatidylcholine (= DMPC), and distearoyl phosphatidylcholine (= DSPC) in aqueous environment. The two variables in this system are temperature, and  the fraction of DSPC. These two variables represent the two axes. In  the gray shaded region a SO and a LD  phases coexist. At a given  temperature, the composition of the two phases are given by the lines in the phase diagram (for details see  \cite{Lee1977b}). Above the upper line (liquidus line), the whole membrane is in the LD (fluid) phase, while it is in the SO (gel) phase below the lower 
line (solidus line).\\

Biological membranes, however, consist of hundreds of components, which are the different lipids, the membrane proteins, and the many small molecule components.  What is immediately clear from the above is that in a biological membrane with several 100 components one may also have many different phases coexisting.  For $K=100$ one may in principle find up to 101 coexisting phases ($P=101$).  Thus, the thermodynamic behavior of such membranes may be very complex, and it may be practically impossible to unravel the detailed behavior of such complex mixtures.  Nevertheless, clearly such phase coexistence is present in biological membranes and must strongly influence their behavior. Therefore, to make general statements on the  physical behavior of biomembranes, one may wish to find general relations that do not require the detailed knowledge of the phase diagram. Such relations exist and will be outlined in section \ref{sec5}.

\subsection{Domain formation}
\label{sec4.3}
The concept of a phase is a strictly theoretical one assuming, for instance, that the system under consideration is infinitely big, and that the phases are separated from each other (i.e., their interface can be neglected).  Both assumption are not quite true for biological membranes.  Cells are relatively small (on the order of $10^{9}$ lipids) and phases cannot be macroscopically separated.   Instead, one typically obtains finite size domains that interact at their boundaries.  What can be concluded from the above is that both artificial membranes made of lipid mixtures, and biological membranes must display the coexistence of many domains of different physical properties.
\begin{figure}[H]
    \hspace*{0.0cm}
    \includegraphics[width=7.5cm]{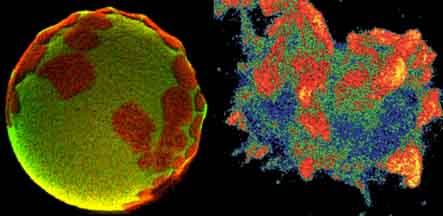}\\
    \hspace*{0.5cm}\parbox[c]{7cm}{\caption{\textit{Domain formation.
    Left: On a vesicle made of a lipid mixture of DLPC and DPPC
    (courtesy M. Fidorra, NBI).  Right: On a living fibroblast cell
    (adapted from \cite{Gaus2003}).  The dimension of both object is about 30
    $\mu$m.}
    \label{Figure4C}}}
\end{figure}

\subsubsection{Domain formation in artificial membranes and monolayers}
\label{sec4.3.1}
Domain formation in artificial systems is actually quite easy to observe.  Already in the early 1980's domains were described in monolayer films (e.g., \cite{Weis1984, Mohwald1990}).  The first demonstration of domain formation on bilayer membranes was from \cite{Korlach1999} in 1999.  In Fig.  \ref{Figure4C} (left) we show a confocal fluorescence microscopy image of a giant vesicle ({\O} $\approx 30 \mu$m) made of a lipid mixture of dilauroyl phosphatidylcholine (DLPC) and dipalmitoyl phosphatidylcholine (DPPC) and measured at room temperature.  According to the phase diagram of this lipid mixture it should display phase coexistence at room temperature.  This can be seen as red and green domains that have been labelled with two different fluorescence markers with affinity for the two different phases.  Red domains are in the SO phase, while the green regions are LD phases.  Such domains are obviously rather big, i.e. of order of 5-10 $\mu$m.

\begin{figure}[H]
    \hspace*{1.5cm}
    \includegraphics[width=5cm]{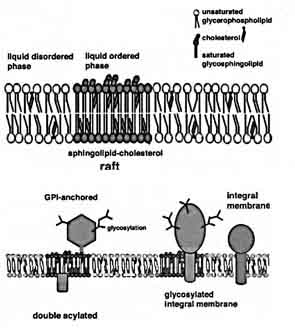}\\
    \hspace*{0.5cm}\parbox[c]{7cm}{\caption{\textit{The formation of 
    cholesterol-rich rafts with GPI-anchored proteins as envisioned 
    by \cite{Bagnat2002}.}
    \label{Figure4D}}}
\end{figure}

\subsubsection{Domain formation in biological systems}
\label{sec4.3.2}
Domain formation in biological membranes is much more difficult to monitor.  In the 1990's, a number of authors proposed the existence of domains called 'rafts' \cite[]{Simons1997, Brown1998, London2002, Edidin2003a, Edidin2003b}. They are thought to be rich in cholesterol, sphingolipids and some GPI-anchored proteins that are in the LO phase (Fig. \ref{Figure4D}).  The physical evidence for such domains in real biomembranes is relatively weak.  If they exist, they seem to be smaller than the resolution of a light microscope (< 200 nm).  Further, due to the many different components of biological membranes there is no need to constrain the concept of domain formation to particular domain compositions. There may be domains of all kinds of compositions, e.g. protein clusters (already shown in the famous paper by \cite{Singer1972}), but also SO and LD domains. One of the nicest demonstrations of domains in cells to date is from \cite{Gaus2003} in fibroblast cells shown in Fig. \ref{Figure4C} (right).

\section{Curvature and elastic constants}
\label{sec5}
Membranes are typically not flat rigid sheets but often display curvature when they form vesicles and organelles. They are soft and display both fluctuations in area and curvature. Further, when membranes are asymmetric they may curve spontaneously. 

\begin{figure}[H]
    \hspace*{2.0cm}
    \includegraphics[width=4cm]{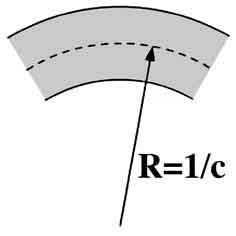}\\
    \hspace*{0.5cm}\parbox[c]{7cm}{\caption{\textit{Curvature and 
    radius of curvature for a bend membrane.}
    \label{Figure5A}}}
\end{figure}
\subsection{Principle curvatures}
\label{sec5.1}
Bending can be characterized by its radius of curvature, $R$, or its curvature, $c=1/R$ (Fig. \ref{Figure5A}). Since the surface has two dimensions one typically deals with two principle curvatures, $c_{1}$ and $c_{2}$ that may be different.\\ 

To curve the membrane one has to provide free energy. The free energy density of a membrane, $g$, is given by a famous formula by \cite{Helfrich1973}:
\begin{equation}
    g=\frac{1}{2}K_{B}\left(c_{1}+c_{2}-c_{0}\right)^2+K_{G}c_{1}c_{2}    
    \label{eq:5.1}
\end{equation}
This equation contains three material constants that have to be determined experimentally: $K_{B}$ is the bending modulus, $K_{G}$ is the Gaussian curvature modulus, and $c_{0}$ is the spontaneous curvature. The spontaneous curvature indicates the equilibrium curvature of the membrane. It mainly depends on the asymmetry of the membrane, for example, different lipid and protein compositions in the two opposing monolayers of the membrane, but also electrostatic potential differences, pH differences, etc. 

\subsection{Curved phases}
\label{sec5.2}
\begin{figure}[H]
    \hspace*{0cm}
    \includegraphics[width=7.5cm]{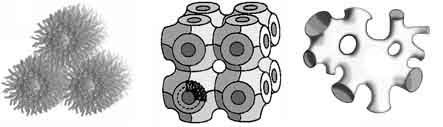}\\
    \hspace*{0.5cm}\parbox[c]{7cm}{\caption{\textit{Lipid phases with 
    different curvatures. Left: Inverse hexagonal phases (from 
    \cite{Heimburg2007a}), Center: Cubic phase 
    (adapted from \cite{Lindblom1989}). Right: Sponge phase (from 
    \cite{Heimburg2007a}). }
    \label{Figure5B}}}
\end{figure}

\begin{figure}[H]
    \hspace*{1.0cm}
    \includegraphics[width=5.5cm]{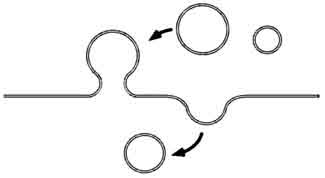}\\
    \hspace*{0.5cm}\parbox[c]{7cm}{\caption{\textit{Fusion and fission events as 
    in endocytosis and exocytosis (adapted from \cite{Heimburg2003}) that 
    involve changes in local curvature.}
    \label{Figure5C}}}
\end{figure}

There exist various curved lipid phases. \\
The inverse hexagonal phase (Fig.  \ref{Figure5B}, left) is not a membrane phase.  It forms due to the spontaneous curvature of the monolayers. There has been quite some discussion of whether the potency of biological membranes to form such phases may be linked to biological function \cite[]{Toombes2002}.  \\
The cubic phase (Fig.  \ref{Figure5B}, center) resembles a periodic three-dimensional lattice of vesicles that are connected. It is a bicontinuous bilayer structure with no inside and outside. They display saddle-point structures.  These are the geometries where the two principle curvatures, $c_{1}$ and $c_{2}$ have opposite sign.\\
The sponge phase (Fig.  \ref{Figure5B}, right) has the same topology than the cubic phase, but it does not display any periodic regularity.\\
Fig.  \ref{Figure5C} shows the biological processes of endocytosis and exocytosis, i.e. fusion and fission of vesicles and membranes. They obviously display the a very similar saddle-point  geometry at the site of fusion as the cubic and the sponge phase. This triggered the interest in such membrane phases \cite[]{Lohner1996, Toombes2002}.

\subsection{Elastic constants}
\label{sec5.3}
The volume compressibility is the function that indicates how much the volume changes when the pressure is increased by an incremental amount. Very compressible substances display a large change in volume when the pressure increases. The area compressibility is the respective function that indicates how much the area changes upon changes in lateral pressure or surface tension. \\
The curvature elasticity is the inverse of the curvature modulus as defined in eq. \ref{eq:5.1}. \cite{Evans1974} has shown that the curvature elasticity is linearly related to the lateral compressibility.

\subsection{Temperature and pressure dependence of elastic constants}
\label{sec5.4}
The elastic constants are actually not really constants. They themselves depend in a sensitive manner on the thermodynamic variables, for example on temperature, pressure, etc. In particular, it has been shown by \cite{Heimburg1998} and \cite{Ebel2001} that the elastic constants all depend in a simple manner on the heat capacity of the membrane that assumes high values in the transitions of membranes (see Figs.  \ref{Figure3A} and \ref{Figure3B}).  This means that within transitions, membranes are very elastic.  They are soft, easy to bend and easy to compress.  Since biological membranes display such transitions close to body or growth temperature, this is of quite high significance.  Everything that is related to the elastic constants will respond very sensitively to changes in temperature and other variables like pH, calcium content etc.  This includes membrane fusion events as well as the spontaneous formation of pores in the membrane as shown in section \ref{sec7}.  We further show in section \ref{sec8} that close to such transitions one obtains the possibility to propagate stable pulses along cylindrical membranes.

\section{Electrostatic potential}
\label{sec6}
On average about 10\% of all lipids are negatively charged.  As mentioned in section \ref{sec2} the relative amount of negative charge is very different depending on the particular organelle of a cell. Especially charged are the membranes of mitochondria that contain up to 20\% cardiolipin that have two negative charges each. Further, the charges often distribute asymmetrically such that the inner membranes of cells and organelles are more charged \cite[]{Rothman1977, Rothman1977b}.\\

\subsection{Ionic strength dependence}
\label{sec6.1}
\begin{figure}[H]
    \hspace*{0.0cm}
    \includegraphics[width=7.5cm]{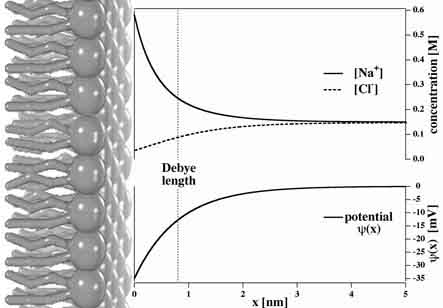}\\
    \hspace*{0.5cm}\parbox[c]{7cm}{\caption{\textit{Distribution of
    positive (Na$^+$) and negative ions (Cl$^-$), and magnitude of the electrostatic
    potential, $\Psi$, as a function of distance from a membrane containing
    acidic lipids.  From \cite{Heimburg2007a}}
    \label{Figure6A}}}
\end{figure}

The electrostatic theory of surfaces originates from Gouy and Chapman (see, e.g \cite{Jahnig1976}).  It describes how the electrostatic potential on surfaces depends on the charge density and on the ionic strength of the salt solution.  For membranes one finds that the electrostatic potential decays fast as a function of the distance from the membrane. The characteristic decay length is called the Debye length. For a 100mM solution of NaCl it is 0.97 nm. For 10mM NaCl it is about 3.1 nm and for 1mM it is 9.7 nm. Close to the membrane the Na$^+$ concentration is increased, while the Cl$^-$ concentration is reduced (Fig. \ref{Figure6A}).

\subsection{Dependence on charged lipid concentration}
\label{sec6.2}
Naturally, the electrostatic potential at a membrane surface also depends on the number of charged molecules the form the membrane. For the ionic strength conditions that are normally found in biology, the surface potential is just proportional to the charge density on the surface and inversely proportional to the square root of the ionic strength.

\subsection{Dependence on bound proteins}
\label{sec6.3}
Proteins also carry charges. Many soluble proteins are positively charged and therefore interact with negatively charged lipid membranes. One example is cytochrome c, a protein from the respiratory chain in mitochondria, with a net positive charge of about 4. It binds to the highly charged mitochondrial membranes and reduces the overall electrostatic potential.

\section{Permeability}
\label{sec7}
\begin{figure}[H]
    \hspace*{1.0cm}
    \includegraphics[width=6cm]{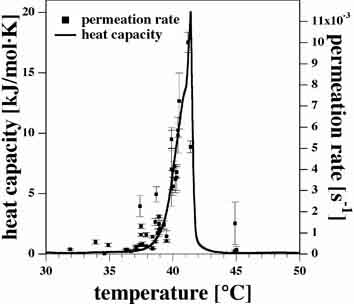}\\
    \hspace*{0.5cm}\parbox[c]{7cm}{\caption{\textit{Permeability of 
    a lipid vesicles for a fluorescent dye as a function of temperature compared to the 
    melting profile \cite[]{Blicher2009}.}
    \label{Figure7A}}}
\end{figure}

Many biology textbooks state that lipid membranes are impermeable to water, ions and small molecules. Although this seems intuitive because the inner core of a membrane is made of hydrocarbons, this is not quite correct. The lipid layers are very thin and due to fluctuations in molecular there is always some finite likelihood that spontaneous pores are formed. This likelihood is dramatically increased in the phase transition temperature regime of lipid membranes, i.e., slightly below body temperature for biomembranes \cite[]{Papahadjopoulos1973, Nagle1978b}. Fig. \ref{Figure7A} shows the permeability of artificial membranes for a fluorescence dye as a function of temperature compared to the heat capacity. One can recognize that the permeability is closely related to the heat capacity. 

\subsection{Protein ion channels}
\label{sec7.1}
\begin{figure}[H]
    \hspace*{0.0cm}
    \includegraphics[width=7.5cm]{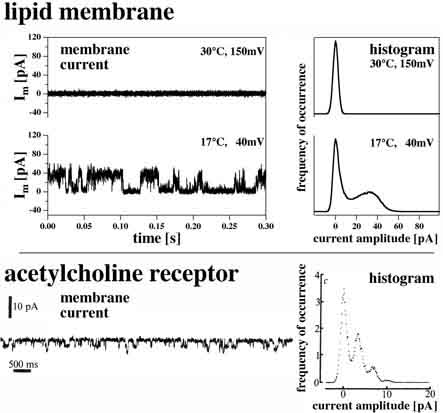}\\
    \hspace*{0.5cm}\parbox[c]{7cm}{\caption{\textit{Microscopic permeability of membranes in the presence and absence of proteins. Bottom: Spontaneous ion currents through a biological membrane in the presence of the acetylcholine receptor adapted from the original paper of \cite{Neher1976}. Top: Spontaneous ion currents through a synthetic black lipid membrane at its phase transition (17$^{\circ}$C).  Above its transition at 30$^{\circ}$C the currents disappear \cite[]{Blicher2009}.}
\label{Figure7B}}}
\end{figure}

Since the permeability of lipid membranes is thought to be small, most of the permeation events through membranes have been attributed to proteins that form pores or ion channels. In patch clamp or black lipid membrane (BLM) measurements on finds quantized ion currents through the membrane indicated localized events. Fig. \ref{Figure7B}(bottom) shows the first of such traces recorded in the presence of the acetylcholine receptor protein published by \cite{Neher1976}. Both scientists received the Nobel prize in 1991. The currents occur in discrete steps that can be seen in the histogram at the right hand side. For a detailed description of ion channel proteins see \cite{Hille1992}.

\subsection{Lipid ion channels}
\label{sec7.2}
As already seen in Fig. \ref{Figure7A} the lipid membranes are also permeable to molecules if they are close to their phase transition temperature. In patch clamp or BLM measurements one finds quantized ion currents through the membrane. This can be seen in Fig. \ref{Figure7B} (top) for a lipid mixture with transition maximum at 17$^{\circ}$C. The currents at the transition temperature resemble those seen in the biological membrane describe above. At 30$^{\circ}$C the currents disappear. This effect is not as well known as the protein conductance but has been described by a number of authors in the literature (e.g., \cite{Antonov1980, Kaufmann1983a, Kaufmann1983b, Gogelein1984, Antonov2005}. Protein ion channels are always embedded in membranes. Biological membranes are close to their transition temperature where they can also show quantized channels due to events in the lipid membrane. Therefore, it does not seem straight-forward how to distinguish between the two events.

\section{Pulse propagation}
\label{sec8}
Nerve cells form long membrane cylinders that are called dendrites or axons. Along these cables one finds the propagation of voltage pulses that are called action potentials. However, also mechanical pulses can travel along nerve membranes. It turns out that these two seemingly different events may well be just be caused by one single phenomenon. 

\subsection{Nerve pulses}
\label{sec8.1}
Action potentials have been described by \cite{Hodgkin1952} (Nobel prize 1963). Their theory is based on the existence of voltage-gated ion channels (in particular of sodium and potassium channels, see Fig. \ref{Figure2B}). It is an electrical theory in which the lipid membrane serves as a capacitor and the ion channel proteins are electrical resistors. The resulting differential equation has the shape of a propagating voltage pulse very similar to that observe in real nerves. Since it is a purely electrical model it does not contain and describe any changes in other variables, e.g. heat, volume, thickness and area.

\subsection{Soliton propagation}
\label{sec8.2}
Biological membranes are slightly above their transition temperature where both the elastic constants and the permeability change drastically. \cite{Heimburg2005c} showed that under these conditions stable mechanical pulses (called solitons) can travel along membranes. Since membranes possess both dipole potentials and electrostatic potentials, these mechanical pulses are accompanied by changes in voltage. The properties of such pulses are based on reversible physics. Area, thickness and voltage of the membranes change during the pulses. Since the events are reversible, the process is isentropic, i.e., no heat is dissipated during the pulse. During the pulse all variables change that are associated to the phase transition in membranes. The free energy necessary to generate a pulse is given by its distance to the phase transitilidocaineon.

\subsection{Reversible heat}
\label{sec8.3}
A problem for the Hodgkin-Huxley model is the experimental fact that no heat is dissipated during the nerve pulse \cite[]{Abbott1958, Howarth1968, Ritchie1985}. Since currents flow through resistors (the ion channels), heat production would be expected during the action potential. Instead, one first finds a phase of heat liberation followed by a phase of heat reabsorption. This finding is rather in agreement with a reversible electromechanical pulse as described by the soliton theory.

\subsection{Mechanical changes}
\label{sec8.4}
Nerves change their dimensions during the action potential. Tasaki and coworkers found that during the action potential the nerves change their thickness by about 1 nm (e.g., \cite{Iwasa1980a, Iwasa1980b, Tasaki1990} and that nerves also get shorter upon excitation \cite[]{Tasaki1989}. Thus, additional to the electrical phenomena mechanical changes also exist during nerve pulses.

\section{Anesthesia}
\label{sec9}
Anesthesia is a phenomenon caused by a number of small molecules that partition in the biological membrane. For more than hundred years it is known that the effectiveness of anesthetics is proportional to their solubility in olive oil (that has the properties of the membrane interior). This rule is known as the Meyer-Overton rule \cite[]{Overton1901}. It holds over several orders of magnitude ranging from laughing gas, N$_2	$O, over halothane to lidocaine. Even the noble gas xenon is an anesthetic. This excludes any specific binding to macromolecules (e.g., proteins) if one is searching for a generic explanation of anesthesia. It has also been known for a long time that anesthetics cause a lowering of phase transition temperatures \cite[]{Trudell1975, Kharakoz2001, Heimburg2007c}. This has already been discussed in section \ref{sec3.1.3}. There are strong indications that the effect of anesthetics is related to this finding. As shown in section \ref{sec3.1.2}, phase transitions are pressure dependent. While pressure increases transitions, anesthetics lower them. It has in fact been found that pressure reverses the effect of anesthesia \cite[]{Johnson1950}.  In the context of the soliton theory (section \ref{sec8.2}), anesthetics increase the free energy necessary to generate an electromechanical pulse \cite[]{Heimburg2007b, Heimburg2007c}. There is no obvious connection between the Hodgkin-Huxley model and anesthesia, but there are attempts to relate the action of anesthesia to specific binding to in channel proteins.

\section*{Acknowledgments}
\label{sec10}
Dr. Manfred Konrad from G\"ottingen prepared the \textit{E.coli} bacteria used to obtain the data shown in Fig. \ref{Figure3B}.
\end{multicols}

\begin{tabular}{p{16cm}}
    \\ \hline\\
\end{tabular}

\begin{multicols}{2}[]

\small{

}
\end{multicols}

\end{document}